# DIELECTRIC AND REFRACTIVE INDEX MEASUREMENTS FOR THE SYSTEMS 1-PENTANOL + OCTANE, OR + DIBUTYL ETHER OR FOR DIBUTYL ETHER + OCTANE AT DIFFERENT TEMPERATURES.

VÍCTOR ALONSO, JUAN ANTONIO GONZÁLEZ*, ISAÍAS GARCÍA DE LA FUENTE, JOSÉ CARLOS COBOS

G.E.T.E.F., Departamento de Física Aplicada, Facultad de Ciencias, Universidad de Valladolid, 47071 Valladolid, Spain,
*e-mail: jagl@termo.uva.es; Fax: +34-983-423136; Tel: +34-983-423757

**ABSTRACT**

Relative permittivities, $\varepsilon_r$, and refractive indixes, $n_D$, have been measured at (288.15-308.15) K and (293.15-303.15) K, respectively, for the mixtures 1-pentanol + octane, or + dibutyl ether and dibutyl ether + octane. These data have been used, together with density measurements available in the literature, to determine the correlation factor, $g_K$, for the studied systems according to the Kirkwood-Fröhlich equations. Results show that the existence of cyclic species of 1-pentanol are predominant at low concentrations of this alkanol when is mixed with octane. These species are broken in large extent by the more active molecules of oxaalkane in the dibutyl ether mixtures, which is in agreement with calorimetric data. The dibutyl ether + octane system does not show meaningful structure. These conclusions are confirmed by values of the molar polarization and by the temperature dependence of $\varepsilon_r$. The empirical expressions of Lorentz-Lorenz, Wiener, Heller, Gladstone-Dale and Newton correlate well the $n_D$ data.

## 1. INTRODUCTION

Interactions and structural effects in liquid mixtures of non-electrolytes have a marked effect on their thermodynamic properties such as phase equilibria, or excess molar functions, enthalpies ($H_m^E$), isobaric heat capacities ($C_{pm}^E$) or volumes ($V_m^E$). This type of data may be used to investigate orientational and structural effects by means of the application of different theories. The Flory model [1] allows the characterization of deviations from random mixing by studying the concentration dependence of the interaction parameter, $X_{12}$ [2-4]. The Kirkwood-Buff formalism [5,6] is concerned with the study of fluctuations in the number of molecules of each mixture component, and of crossed fluctuations [3,4]. The ERAS model [7] is a useful tool to characterize self-association and solvation effects [8]. The purely physical model DISQUAC [9,10] can describe accurately a whole set of thermodynamic properties such those mentioned above (except $V_m^E$, as it is a rigid lattice model and $V_m^E = 0$ it is assumed) [8,11], the Kirkwood-Buff integrals [11,12], or concentration-concentration structure factor [13]. Other physical properties which are also important to gain insight into interactional effects and those related to molecular size and shape are viscosity [14,15], refractive index [16,17] or permittivity. Measurements of this magnitude together with $n_D$ and density data can be used to determine the Kirkwood correlation factor, $g_K$, [18-21] which provides useful information on the mixture structure (see below) [18-21]. In this work, we report $\varepsilon_r$ data at (288.15-308.15) K for the systems 1-pentanol + octane, + dibutyl ether, or for dibutyl ether + octane. Data on $n_D$ at (293.15-303.15) K are also given for the mentioned systems. These measurements together with those available in the literature for the density of the systems [22-24] are used to calculate $g_K$. Derived quantities as polarizability volume [25], or molar refraction [26,27] are also considered to a better understanding of the interactional and structural effects present in the investigated solutions. Data on $\varepsilon_r$ and $n_D$ 298.15 K for the 1-pentanol systems are already available in the literature [28-30]. The $n_D$ measurements have been correlated using the following empirical equations: Lorentz-Lorenz, Weiner, Heller, Gladstone-Dale and Newton [31,32].

## 2. EXPERIMENTAL

### 2.1 Materials

1-Pentanol, and octane were supplied by Fluka, and dibutyl ether by Sigma-Aldrich and were used without further purification. Their purity in mass fractions was ≥ 0.99; ≥ 0.99 and ≥ 0.995, respectively. Values of the physical properties of pure compounds, density, $\rho$, (measured with an Anton Paar DMA 602 vibrating-tube densimeter; uncertainty 5 g·cm$^{-3}$) relative permittivity, $\varepsilon_r$, and refractive indixes, $n_D$, are listed in Table 1. They are in good agreement with the literature values (Table 1).

*2.2 Apparatus and procedures*

Binary mixtures were prepared by mass in small flasks of about 10 cm$^3$. All weighings were corrected for buoyancy effects. The error on the final mole fraction is estimated to be lower than ± 0.0001. Conversion to molar quantities was based on the relative atomic mass table of 1995 issued by I.U.P.A.C [33]. All the measurements of the thermophysical properties were carried out under static conditions and atmospheric pressure

Permittivity measurements were carried out using the Agilent 16452A cell, connected to a precision impedance analyzer model 4294A through a 16048G test lead, both also from Agilent. The frequency range of the impedance analyzer is from 40 Hz to 110 MHz, and that for the measuring cell is from 20 Hz to 30 MHz. Temperature is controlled within ± 0.02 K by means of a thermostatic bath LAUDA RE304. Different spaces for the cell can be selected in order to vary the separation between electrodes and, consequently the volume of the dielectric under study. The calibration procedure led to use a separation of 3 mm, which corresponds to a total volume of the dielectric of 6.8 cm$^3$. A scheme of the measuring cell and of the experimental arrangement are shown in Figure 1. Measurements were taken, for all the samples, at the frequencies: (1, 10, 100) kHz and (1, 10) MHz. Calculations were carried out at 1 MHz, as the results are less scattered. In addition, according to specifications from the manufacturer, the relative error is smaller. This configuration leads to an estimated error in the electric capacity of the cell of 0.1% and to an accuracy for the relative permittivity of 3% or less. Calibration was developed with the following pure liquids: water, benzene, cyclohexane, hexane, nonane, decane, dimethyl carbonate, diethyl carbonate, methanol, 1-propanol, 1-pentanol, 1-hexanol, 1-heptanol, 1-octanol, 1-nonanol and 1-decanol in the temperature range (288.15-333.15) K. From the differences between our data and values available in the literature, the uncertainty of $\varepsilon_r$ is estimated to be l% or less.

Refractive indices were measured with a refractometer model RMF970 from Bellingham-Stanley. The accuracy of the apparatus is ± 0.00002. The measurement method is based on the optical detection of the critical angle at the wave length of the sodium D line (586.9 nm). The temperature is controlled by Peltier modulus and the temperature stability is ± 0.05 K. Prior to the measurement at each temperature, the apparatus must be calibrated with a known reference, usually distilled and deionised water or toluene. The uncertainty of the $n_D$ measurements is 0.02% or better, as it is shown by the differences between our $n_D$ results and those reported in the literature for recommended liquids [34] (methylcyclohexane, *iso*-octane, octane and hexadecane).

## 3. RESULTS

Table 2 lists, in the temperature range (288.15-308.15) K, values of $\varepsilon_r$ and of deviations of this magnitude from the ideal state vs $x_1$, the mole fraction of the first component for 1-pentanol + octane, or + dibutyl ether, or for dibutyl ether + octane systems. For an ideal mixture at the same temperature and pressure than the system under study, the $\varepsilon_r^{id}$ values are calculated from the equation [35]:

$$\varepsilon_r^{id} = \phi_1 \varepsilon_{r1} + \phi_2 \varepsilon_{r2} \qquad (1)$$

where $\phi_i = x_i V_i / V^{id}$. Deviations from the ideal behaviour are then calculated according to the expression:

$$\Delta \varepsilon_r = \varepsilon_r - \phi_1 \varepsilon_{r1} - \phi_2 \varepsilon_{r2} \qquad (2)$$

Table 3 contains values for $n_D$ at (293.15-303.15) K for the investigated solutions. Values of $\Delta n_D$ (= $n_D - n_D^{id}$) calculated using [36]:

$$n_D^{id} = [\phi_1 n_{D1}^2 + \phi_2 n_{D2}^2]^{1/2} \qquad (3)$$

are not included as the data are close to the experimental uncertainties, and therefore the relative errors are very important. Results on $\Delta \varepsilon_r$ and $n_D$ are shown graphically in Figures 2-3. Our $n_D$ results compare well those available in the literature. The $\Delta \varepsilon_r$ data were fitted by unweighted least-squares polynomial regression to the equation:

$$\Delta \varepsilon_r = x_1 (1 - x_1) \sum_{i=0}^{k-1} A_i (2x_1 - 1)^i \qquad (4)$$

The number of coefficients $k$ used in eq. (4) for each mixture was determined by applying an F-test [37] at the 99.5 % confidence level. Table 4 lists the parameters $A_i$ obtained in the regression. The $n_D(x_1)$ data have been fitted to the function:

$$n_D = \sum_{i=0}^{2} A_i x_1^i \qquad (5)$$

Values of the $A_i$ coefficients are given in Table 4, which also lists the corresponding standard deviations, $\sigma$, defined by:

$$\sigma(F) = \left[\frac{1}{N-k}\sum(F_{cal} - F_{exp})^2\right]^{1/2} \qquad (6)$$

where $N$ is the number of direct experimental values and $F = \Delta\varepsilon_r$ or $n_D$. In addition, the ability of the equations of Lorentz-Lorenz, Wiener, Heller, Gladstone-Dale and Newton [31,32] to describe the $n_D$ measurements has been tested. Expressions for these equations can be found elsewhere [31,32] and will be not repeated here. All the equations provide very similar results for the considered systems. Deviations obtained from the Lorentz-Lorenz equation are 0.04%, 0.05% and 0.02% for the systems 1-pentanol + octane, + dibutyl ether, or dibutyl ether + octane at 293.15 K, respectively. At 298.15 K, these deviations are 0.02%, 0.02% and 0.03%, and, at 303.15 K, they are: 0.02%, 0.02% and 0.04%. It can be concluded that the $n_D$ data are correctly described by the mentioned equations.

## 4. DISCUSSION

We note that $\Delta\varepsilon_r < 0$ for all the investigated mixtures (Table 2, Figure 2). In the case of 1-pentanol solutions, this reveals that octane or dibutyl ether molecules act as structure breakers of the alcohol structure, which leads to a decrease in the total number of parallel aligned effective dipoles of 1-pentanol that contribute to the dielectric polarization of the system [38]. Similarly, the addition of octane to pure dibutyl ether implies the breaking of the much weaker structure of this ether.

The $\varepsilon_r$ and $n_D$ measurements at 298.15 K have been used, together with excess molar volume data available in the literature [22-24], to calculate the correlation factor, $g_k$, of the three binary mixtures investigated. For a system containing one polar compound and one non-polar component, $g_k$ can be determined by the Fröhlich equation [18,19,39,40]

$$g_k = \frac{9k_B T \varepsilon_0 (2\varepsilon_r + \varepsilon_1^\infty)^2}{N_A \mu_1^2 x_1 (\varepsilon_1^\infty + 2)^2 (2\varepsilon_r + 1)}\left[\frac{V_m(\varepsilon_r - 1)}{\varepsilon_r} - \frac{3V_1 x_1(\varepsilon_1^\infty - 1)}{2\varepsilon_r + \varepsilon_1^\infty} - \frac{3V_2 x_2(n_{D2}^2 - 1)}{2\varepsilon_r + n_{D2}^2}\right] \qquad (7)$$

where $k_B, N_A, \varepsilon_0$ and $T$ are the Boltzmann´s constant, the Avogadro´s number, the dielectric constant of the vacuum and the system temperature, respectively; $\mu_1$ is the dipole moment of the polar compound (component 1) (1.65 D [41] for 1-pentanol and 1.18 D for dibutyl ether [41] (1

D = 3.3564 $10^{-30}$ C·m)); $V_m$ and $V_i$ are the molar volumes of the mixture and of the component i, while $\varepsilon_r$ and $\varepsilon_1^\infty$ are, respectively, the measured dielectric constant of the system, and the high frequency dielectric constant of the polar compound 1. The $\varepsilon_1^\infty$ magnitude was calculated from the Clausius-Mossotti equation [21], adopting the atomic polarization ($P_A$) that was evaluated using the relation

$$P_A + P_E = 1.1 P_E \tag{8}$$

where $P_E$, the electronic polarization, was calculated by the Lorenz-Lorentz equation using the refractive index for the sodium-D line [18,19,40]. On the other hand, for a binary mixture involving two polar compounds, $g_k$ is given by [21,25,42]:

$$g_k = \frac{9 k_B T V_m \varepsilon_0 (\varepsilon_r - \varepsilon_r^\infty)(2\varepsilon_r + \varepsilon_r^\infty)}{N_A \mu^2 \varepsilon_r (\varepsilon_r^\infty + 2)^2} \tag{9}$$

Here, $\varepsilon_r^\infty$ is the high frequency dielectric constant of the system, calculated similarly to $\varepsilon_1^\infty$, i.e., using equation (8) [43,44] and $\mu$ represents the gas phase dipole moment of the solution. Due to the lack of experimental data, they have been estimated from the equation [25]:

$$\mu = x_1 \mu_1 + x_2 \mu_2 \tag{10}$$

where $\mu_i$ stands for the dipole moment in the gas phase of component i. Uncertainties of $g_K$ are less than 7% or 4%, when calculating this magnitude from equations (7) and (9), respectively. Figure 4 compares our $g_k$ values for the 1-pentanol + octane mixture with those available in the literature. The good agreement between them is remarkable.

It is known that unstructured solvents are characterized by $g_k = 1$, while structured solvents generally have $g_k > 1$. This is ascribed to the dipoles are aligned in a parallel way. If they are aligned in antiparallel way, then $g_k < 1$ [18,45,46]. The observed $g_k$ minimum in the 1-pentanol + octane mixture (Figure 4) is characteristic of the 1-alkanol + alkane systems [18-20,40,46]. According to the rules of vector addition, the dipole moment of a hydrogen-bonded cyclic group is zero, and the dipole moment of a linear chain is greater than that of a monomer. Thus, the formation of cyclic species leads to a decrease of the apparent dipole moment, and the formation of linear chains has the opposite effect. Therefore, the existence of the $g_K$ minimum reveals a predominance of the cyclic species, mainly tetramers [20], over the linear chains.

It is interesting to compare the $g_k$ curves for the 1-pentanol + octane, or + dibutyl ether systems (Figure 4). We note that at low concentration of the alcohol, $g_k$ (octane) < $g_k$ (dibutyl ether), while the opposite behaviour is encountered at high 1-pentanol concentrations. This is indicative of a decrease of the effects related to the self-association of the alcohol, as ethers are more active breakers of the polymeric species of 1-alkanols than alkanes [3]. In terms of $H_m^E$, this means that for solutions including a given alcohol, at 298.15 K and equimolar composition, $H_m^E$ (dibutyl ether) > $H_m^E$ (octane); e.g., in the case of 1-pentanol mixtures, 843 (dibutyl ether) [47] > 614 (octane) [48] (both values in J·mol$^{-1}$). Moreover, $H_m^E$ curves are more skewed to higher mole fractions of the 1-alkanol in solutions with ethers, indicating that effects related to the self-association of the alcohols are less important [3]. As a consequence, $g_k$ (dibutyl ether) > 1 at low alkanol concentration and now the cyclic polymers are not predominant in that region. We also note that $g_K$ (octane) > $g_K$ (dibutyl ether) at higher alkanol concentrations, as ethers also break more easily than alkanes the linear polymeric chains of 1-pentanol.

The temperature dependence of $g_k$ for the 1-pentanol solutions has been examined using density data given elsewhere [23,49]. For the 1-pentanol + octane system, $g_k$ increases with the temperature in the dilute alcohol region, while the opposite trend is observed at high alkanol concentrations. For example, at $x_1$ = 0.05, $(\Delta g_k / \Delta T)_{298.15}$ = 0.026 K$^{-1}$ and at $x_1$ = 0.90, $(\Delta g_k / \Delta T)_{298.15}$ = −0.015 K$^{-1}$. This again supports the existence of cyclic species in dilute region of alcohol. Note that, in terms of the ERAS model, the equilibrium constant for cyclic tetramers is larger than for the linear chains [20]. On the other hand, the temperature dependence of $g_k$ is stronger for the 1-pentanol + octane mixture than for the 1-pentanol + dibutyl ether system. At equimolar composition and 298.15 K, for the former, $\Delta g_k / \Delta T$ = −0.013 K$^{-1}$, while for the latter, this magnitude is −0.006 K$^{-1}$. This is consistent with the higher values of the excess heat capacities at constant pressure, $C_{p,m}^E$, of 1-alkanol + alkane mixtures compared with those of 1-alkanol + ether systems. Thus, at the same conditions that above, $C_{p,m}^E$ /J·mol$^{-1}$·K$^{-1}$ = 11.7, for ethanol + heptane [50], and is 7.2 for ethanol + methyl butyl ether [51].

We have calculated the molar polarization of the mixtures (or polarizability volume) according to the equation [25]:

$$P_{\mathrm{m}} = \frac{(\varepsilon_{\mathrm{r}} - n_{\mathrm{D}}^2)(2\varepsilon_{\mathrm{r}} + n_{\mathrm{D}}^2)V_{\mathrm{m}}}{9\varepsilon_{\mathrm{r}}} \qquad (11)$$

Values of this magnitude at 298.15 K are shown graphically in Figure 5. The main features of $P_{\mathrm{m}}$ are the following. (i) For the octane systems, $P_{\mathrm{m}}$(1-pentanol) $\approx P_{\mathrm{m}}$(dibutyl ether) at low concentrations of the polar compounds. This newly indicates the existence of cyclic species of 1-alkanol in this region which do not contribute to the mixture polarization due to its dipole moment is zero. (ii) At low concentrations of 1-pentanol, $P_{\mathrm{m}}$(dibutyl ether) > $P_{\mathrm{m}}$(octane), which confirms that the ether is a more active molecule when disrupting the cyclic species of the alcohol, in such way that an increase of $P_{\mathrm{m}}$ is produced. (iii) Finally, $P_{\mathrm{m}}$ is similar for the 1-pentanol solutions at high concentrations of this component, that is, when the favourable alignment of the dipoles is predominant.

Molar refractions, $R_{\mathrm{m}}$, (Figure 6) have been calculated using the equation [26,27]:

$$R_{\mathrm{m}} = \frac{(n_{\mathrm{D}}^2 - 1)V_{\mathrm{m}}}{(n_{\mathrm{D}}^2 + 2)} \qquad (12)$$

This equation comes from the Lorentz-Lorenz equation by replacing the refractive index at infinite frequency by the refractive index at optical frequencies (usually, the sodium D line) [21,26,27]. $R_{\mathrm{m}}$ can be regarded as a measure of the dispersion forces within the fluids. It is known that the permanent $\mu$ of polar molecules do not contribute to the polarization of the molecule as the relaxation time of $\mu$ is much larger than the period of oscillation of the light. $R_{\mathrm{m}}$ decreases with increased alkanol concentrations, and this indicates that the polarization of the alkane or dibutyl ether is higher than that of the alkyl chain of the 1-pentanol. Results reveal that $R_{\mathrm{m}}$ is higher for the dibutyl ether + octane mixture than for the systems including 1-pentanol, as in this case orientational polarization is much more important. This is particularly marked at large alcohol concentration. On the other hand, $R_{\mathrm{m}}$ changes linearly with the concentration and does not depend on the temperature [52]. This is the normal behaviour of systems where no complex formation exist [26,52].

We have also calculated the temperature coefficient of the relative permittivity defined as [53]:

$$\alpha = \frac{1}{\varepsilon_{\mathrm{r}}} \frac{d\varepsilon_{\mathrm{r}}}{d(1/T)} \qquad (13)$$

Results (Figure 7) show that $\alpha$ is positive at any concentration, that is, $\varepsilon_r$ decreases when the temperature is increased, which is the normal behaviour of dipolar liquids. On the other hand, it is remarkable that $\alpha$ curves intersect between them at similar concentrations that the corresponding $g_k$ curves (Figures 3,6). This occurs at $x_1 \approx 0.2$ for the 1-pentanol or dibutyl ether + octane mixtures, and at $x_1 \approx 0.4$ for the 1-pentanol + octane, or + dibutyl ether systems. In the case of octane solutions, the fact that $\alpha$ (1-pentanol) < $\alpha$ (dibutyl ether) at $x_1 < 0.2$ also supports the existence of cyclic species of 1-pentanol, which have a weaker response to the application of an external electric field than that of dibutyl ether. At $x_1 > 0.4$, $\alpha$ varies in the sequence: 1-pentanol + octane > 1-pentanol + dibutyl ether > dibutyl ether + octane.

Finally, the temperature dependence of $\varepsilon_r$ and its derivative reflects the behaviour of the electric-field-induced increment of the basic thermodynamic quantities, internal energy, $\Delta U$, entropy, $\Delta S$, and Helmholtz free energy, $\Delta F$. Particularly the entropy variation is given by [53,54]:

$$\frac{\Delta S}{E^2} = \frac{S(T,E) - S_0(T)}{E^2} = \frac{\varepsilon_0}{2} \frac{d\varepsilon_r}{dT} \qquad (14)$$

Where $S_0$ stands for the entropy value in absence of the electric field of amplitude $E$. For the investigated systems, $d\varepsilon_r / dT < 0$ and this means that the application of an electric field to the dielectrics leads to an increase of molecular order when temperature is increased in comparison with case of $E = 0$

## 5. CONCLUSIONS

The properties $\varepsilon_r$ and $n_D$ have been measured at (288.15-308.15) K and (293.15-303.15) K for the systems 1-pentanol + octane, or + dibutyl ether, and dibutyl ether + octane. Values of the correlation factor, calculated at 298.15 K from the Kirkwood-Fröhlich equations, show that dibutylether is a more active compound when breaking the alkanol self-association. This is supported by molar polarization values and by the temperature dependence of $\varepsilon_r$.

**ACKNOWLEDGEMENTS**

The authors gratefully acknowledge the financial support received from the Ministerio de Ciencia e Innovación, under the Project FIS2010-16957. V.A. acknowledges the grant financed jointly by the Junta de Castilla y León and Fondo Social Europeo.

TABLE 1

Properties of pure compounds at temperature $T$: density, $\rho$, dielectric permittivity, $\varepsilon_r$, and refractive index, $n_D$.

| Compound | $T$/K | $\rho$/cm$^3$mol$^{-1}$ | | $\varepsilon_r$ | | $n_D$ | |
|---|---|---|---|---|---|---|---|
| | | Exp. | Lit | Exp. | Lit. | Exp- | Lit |
| 1-pentanol | 288.15 | 0.81849 | 0.8189 [56] | 16.334 | | | |
| | 293.15 | 0.81482 | 0.8147 [57] | 15.736 | | 1.41001 | |
| | 298.15 | 0.81119 | 0.81083 [29] | 15.144 | 15.04 [28] | 1.4084 | 1.40767 [28] |
| | | | 0.810883 [40] | | 15.558 [58] | | 1.4080 [29] |
| | | | 0.81090 [30] | | | | 1.40790 [40] |
| | | | 0.8116 [56] | | | | 1.4077 [30] |
| | | | 0.8107 [58] | | | | 1.4075 [58] |
| | 303.15 | 0.80750 | 0.8073 [57] | 14.565 | | 1.4065 | |
| | 308.15 | 0.80386 | 0.8039 [56] | 13.998 | 13.590 [59] | | |
| Dibutyl Ether | 288.15 | 0.77260 | | 3.186 | | | |
| | 293.15 | 0.76827 | | 3.145 | | 1.3991 | |
| | 298.15 | 0.76397 | 0.76395 [30] | 3.106 | 3.040 [60] | 1.3967 | 1.3965 [30] |
| | | | 0.764067 [60] | | | | 1.3969[60] |
| | 303.15 | 0.75962 | | 3.068 | | 1.3945 | |
| | 308.15 | 0.75531 | 0.755469 [60] | 3.031 | 2.964 [60] | | |
| Octane | 288.15 | 0.70681 | | 1.975 | | | |
| | 293.15 | 0.70277 | 0.70262 [61] | 1.968 | | 1.3978 | |
| | 298.15 | 0.69873 | 0.69850 [29] | 1.961 | 1.96 [28] | 1.3958 | 1.3951 [29] |
| | | | | | 1.95 [62] | | |
| | 303.15 | 0.69470 | | 1.954 | | 1.3930 | |
| | 308.15 | 0.69067 | | 1.947 | | | |

TABLE 2

Dielectric constants, $\varepsilon_r$, at temperature $T$ and deviations from the ideal state, $\Delta\varepsilon_r$, for the systems 1-pentanol(1) + octane(2), or + dibutyl ether(2), or for dibutyl ether(1) + octane(2).

| $x_1$ | $\varepsilon_r$ | $\Delta\varepsilon_r$ | $\varepsilon_r$ | $\Delta\varepsilon_r$ | $\varepsilon_r$ | $\Delta\varepsilon_r$ |
|---|---|---|---|---|---|---|
| | | | 1-pentanol(1) + octane(2) | | | |
| | $T = 288.15$ K | | $T = 293.15$ K | | $T = 298.15$ K | |
| 0.1008 | 2.104 | −0.868 | 2.100 | −0.822 | 2.096 | −0.777 |
| 0.2034 | 2.331 | −1.730 | 2.321 | −1.646 | 2.312 | −1.560 |
| 0.3004 | 2.766 | −2.401 | 2.731 | −2.296 | 2.699 | −2.188 |
| 0.3808 | 3.405 | −2.742 | 3.320 | −2.646 | 3.243 | −2.542 |
| 0.4885 | 4.798 | −2.759 | 4.601 | −2.715 | 4.421 | −2.657 |
| 0.5930 | 6.675 | −2.371 | 6.358 | −2.387 | 6.053 | −2.392 |
| 0.6992 | 8.743 | −1.792 | 8.332 | −1.842 | 7.936 | −1.877 |
| 0.7922 | 11.143 | −1.130 | 10.654 | −1.187 | 10.175 | −1.235 |
| 0.8967 | 13.734 | −0.478 | 13.179 | −0.523 | 12.634 | −0.559 |
| | $T = 303.15$ K | | $T = 308.15$ K | | | |
| 0.1008 | 2.092 | −0.733 | 2.088 | −0.692 | | |
| 0.2034 | 2.304 | −1.476 | 2.297 | −1.394 | | |
| 0.3004 | 2.669 | −2.080 | 2.643 | −1.972 | | |
| 0.3808 | 3.173 | −2.436 | 3.111 | −2.324 | | |
| 0.4885 | 4.257 | −2.587 | 4.109 | −2.506 | | |
| 0.5930 | 5.774 | −2.377 | 5.516 | −2.348 | | |
| 0.6992 | 7.559 | −1.901 | 7.204 | −1.910 | | |
| 0.7922 | 9.712 | −1.277 | 9.267 | −1.307 | | |
| 0.8967 | 12.103 | −0.592 | 11.585 | −0.621 | | |
| | | | 1-pentanol(1) + dibutyl ether(2) | | | |
| | $T = 288.15$ K | | $T = 293.15$ K | | $T = 298.15$ K | |
| 0.1104 | 3.610 | −0.542 | 3.556 | −0.514 | 3.502 | −0.487 |
| 0.1877 | 4.008 | −0.869 | 3.939 | −0.825 | 3.871 | −0.781 |
| 0.3047 | 4.628 | −1.434 | 4.528 | −1.370 | 4.431 | −1.305 |
| 0.4064 | 5.426 | −1.762 | 5.285 | −1.690 | 5.147 | −1.618 |
| 0.5075 | 6.482 | −1.926 | 6.281 | −1.861 | 6.087 | −1.794 |
| 0.6034 | 7.776 | −1.894 | 7.505 | −1.845 | 7.242 | −1.794 |
| 0.7038 | 9.464 | −1.653 | 9.112 | −1.623 | 8.768 | −1.591 |
| 0.8018 | 11.446 | −1.225 | 11.011 | −1.213 | 10.583 | −1.201 |
| 0.9048 | 13.846 | −0.635 | 13.328 | −0.631 | 12.812 | −0.631 |

TABLE 2 (continued)

|  | $T$ = 303.15 K | | $T$ = 308.15 K | | | |
|---|---|---|---|---|---|---|
| 0.1104 | 3.449 | −0.461 | 3.399 | −0.436 | | |
| 0.1877 | 3.805 | −0.738 | 3.741 | −0.696 | | |
| 0.3047 | 4.338 | −1.240 | 4.249 | −1.175 | | |
| 0.4064 | 5.015 | −1.544 | 5.018 | −1.433 | | |
| 0.5075 | 5.901 | −1.723 | 5.725 | −1.653 | | |
| 0.6034 | 6.992 | −1.736 | 6.752 | −1.678 | | |
| 0.7038 | 8.437 | −1.556 | 8.123 | −1.515 | | |
| 0.8018 | 10.169 | −1.185 | 9.771 | −1.166 | | |
| 0.9048 | 12.313 | −0.626 | 11.828 | −0.623 | | |

dibutyl ether(1) + octane(2)

|  | $T$ = 288.15 K | | $T$ = 293.15 K | | $T$ = 298.15 K | |
|---|---|---|---|---|---|---|
| 0.1014 | 2.074 | −0.029 | 2.065 | −0.027 | 2.056 | −0.025 |
| 0.2033 | 2.180 | −0.050 | 2.168 | −0.047 | 2.155 | −0.047 |
| 0.3060 | 2.292 | −0.064 | 2.278 | −0.061 | 2.263 | −0.059 |
| 0.4060 | 2.405 | −0.074 | 2.388 | −0.070 | 2.371 | −0.066 |
| 0.5063 | 2.524 | −0.077 | 2.504 | −0.072 | 2.484 | −0.069 |
| 0.6002 | 2.621 | −0.073 | 2.617 | −0.069 | 2.594 | −0.066 |
| 0.7036 | 2.775 | −0.063 | 2.747 | −0.060 | 2.719 | −0.057 |
| 0.8029 | 2.908 | −0.047 | 2.876 | −0.045 | 2.844 | −0.044 |
| 0.9003 | 3.043 | −0.027 | 3.007 | −0.026 | 2.972 | −0.024 |

|  | $T$ = 303.15 K | | $T$ = 308.15 K | | | |
|---|---|---|---|---|---|---|
| 0.1014 | 2.047 | −0.024 | 2.038 | −0.023 | | |
| 0.2033 | 2.144 | −0.044 | 2.133 | −0.042 | | |
| 0.3060 | 2.249 | −0.056 | 2.235 | −0.053 | | |
| 0.4060 | 2.353 | −0.065 | 2.337 | −0.061 | | |
| 0.5063 | 2.464 | −0.066 | 2.444 | −0.063 | | |
| 0.6002 | 2.570 | −0.064 | 2.548 | −0.061 | | |
| 0.7036 | 2.692 | −0.055 | 2.667 | −0.052 | | |
| 0.8029 | 2.814 | −0.042 | 2.785 | −0.039 | | |
| 0.9003 | 2.938 | −0.023 | 2.905 | −0.022 | | |

TABLE 3

Refractive indexes, $n_D$, at temperature $T$ for the systems 1-pentanol(1) + octane(2), or + dibutyl ether(2), or for dibutyl ether(1) + octane(2).

| $x_1$ | $n_D$ | | |
|---|---|---|---|
| | $T$= 293.15 K | $T$= 298.15 K | $T$= 303.15 K |
| 1-pentanol(1) + octane(2) | | | |
| 0.1008 | 1.3981 | 1.3965 | 1.3937 |
| 0.2034 | 1.3988 | 1.3973 | 1.3946 |
| 0.3004 | 1.3997 | 1.3982 | 1.3955 |
| 0.3808 | 1.4005 | 1.3991 | 1.3964 |
| 0.4885 | 1.4019 | 1.4004 | 1.3979 |
| 0.5930 | 1.4033 | 1.4019 | 1.3993 |
| 0.6992 | 1.4047 | 1.4034 | 1.4008 |
| 0.7922 | 1.4063 | 1.4050 | 1.4026 |
| 0.8967 | 1.4082 | 1.4070 | 1.4046 |
| 1-pentanol(1) + dibutylether(2) | | | |
| 0.1104 | 1.4001 | 1.3976 | 1.3958 |
| 0.1877 | 1.4011 | 1.3985 | 1.3968 |
| 0.3047 | 1.4021 | 1.3995 | 1.3979 |
| 0.4064 | 1.4031 | 1.4008 | 1.3989 |
| 0.5075 | 1.4042 | 1.4017 | 1.4001 |
| 0.6034 | 1.4053 | 1.4029 | 1.4013 |
| 0.7038 | 1.4065 | 1.4041 | 1.4026 |
| 0.8018 | 1.4077 | 1.4054 | 1.4038 |
| 0.9048 | 1.4089 | 1.4067 | 1.4052 |
| dibutyl ether(1) + octane(2) | | | |
| 0.1014 | 1.3979 | 1.3955 | 1.3929 |
| 0.2033 | 1.3979 | 1.3955 | 1.3929 |
| 0.3060 | 1.3980 | 1.3956 | 1.3930 |
| 0.4060 | 1.3981 | 1.3957 | 1.3931 |
| 0.5063 | 1.3982 | 1.3959 | 1.3932 |
| 0.6002 | 1.3983 | 1.3960 | 1.3934 |
| 0.7036 | 1.3985 | 1.3962 | 1.3935 |
| 0.8029 | 1.3987 | 1.3964 | 1.3937 |
| 0.9003 | 1.3989 | 1.3966 | 1.3939 |

TABLE 4

Coefficients $A_i$ and standard deviations, $\sigma$ (eq. 6) for representation of the $\Delta\varepsilon_r$ and $n_D$ properties at temperature $T$ for 1-pentanol(1) + octane(2), or + dibutyl ether(2) or dibuty ether(1) + octane(2) systems by eq. 4 ($\Delta\varepsilon_r$) or by eq. 5 ($n_D$)

| Property | $A_0$ | $A_1$ | $A_2$ | $A_3$ | $\sigma$ |
|---|---|---|---|---|---|
| 1-Pentanol(1) + octane(2) $T$ = 288.15 K | | | | | |
| $\Delta\varepsilon_r$ (eq. 4) | −10.86 | 3.54 | 5.85 | | 0.056 |
| 1-Pentanol(1) + octane(2) $T$ = 293.15 K | | | | | |
| $\Delta\varepsilon_r$ (eq. 4) | −10.70 | 2.73 | 5.65 | | 0.044 |
| $n_D$ (eq. 5) | 1.397 | 0.0054 | 0.0073 | | 0.0006 |
| 1-Pentanol(1) + octane(2) $T$ = 298.15 K | | | | | |
| $\Delta\varepsilon_r$ (eq. 4) | −10.50 | 1.96 | 5.41 | | 0.032 |
| $n_D$ (eq. 5) | 1.396 | 0.0055 | 0.0077 | | 0.0001 |
| 1-Pentanol(1) + octane(2) $T$ = 303.15 K | | | | | |
| $\Delta\varepsilon_r$ (eq. 4) | −10.25 | 1.26 | 5.09 | | 0.024 |
| $n_D$ (eq. 5) | 1.393 | 0.0058 | 0.0078 | | 0.0002 |
| 1-Pentanol(1) + octane(2) $T$ = 308.15 K | | | | | |
| $\Delta\varepsilon_r$ (eq. 4) | −9.95 | 0.63 | 4.71 | | 0.019 |
| 1-pentanol(1) + dibutyl ether(2); $T$ = 288.15 K | | | | | |
| $\Delta\varepsilon_r$ (eq. 4) | −7.68 | −1.47 | 2.23 | | 0.019 |
| 1-pentanol(1) + dibutyl ether(2); $T$ = 293.15 K | | | | | |
| $\Delta\varepsilon_r$ (eq. 4) | −7.42 | −1.66 | 2.05 | | 0.019 |
| $n_D$ (eq. 5) | 1.399 | 0.0087 | 0.0023 | | 0.0003 |
| 1-pentanol(1) + dibutyl ether(2); $T$ = 298.15 K | | | | | |
| $\Delta\varepsilon_r$ (eq. 4) | −7.15 | −1.84 | 1.83 | | 0.019 |
| $n_D$ (eq. 5) | 1.397 | 0.0090 | 0.0024 | | 0.0005 |
| 1-pentanol(1) + dibutyl ether(2); $T$ = 303.15 K | | | | | |
| $\Delta\varepsilon_r$ (eq. 4) | −6.87 | −1.99 | 1.61 | | 0.019 |
| $n_D$ (eq. 5) | 1.395 | 0.0091 | 0.0025 | | 0.0001 |
| 1-pentanol(1) + dibutyl ether(2); $T$ = 308.15 K | | | | | |
| $\Delta\varepsilon_r$ (eq. 4) | −6.44 | −2.25 | 1.02 | | 0.052 |

TABLE 4 (continued)

Dibutyl ether(1) + octane; $T = 288.15$ K

| | | | | | |
|---|---|---|---|---|---|
| $\Delta\varepsilon_r$ (eq. 4) | −0.3048 | | | | 0.0008 |

Dibutyl ether(1) + octane; $T = 293.15$ K

| | | | | | |
|---|---|---|---|---|---|
| $\Delta\varepsilon_r$ (eq. 4) | −0.2875 | 0.0066 | | | 0.0004 |
| $n_D$ (eq. 5) | 1.398 | 0 | 0.0013 | | 0.0002 |

Dibutyl ether(1) + octane; $T = 298.15$ K

| | | | | | |
|---|---|---|---|---|---|
| $\Delta\varepsilon_r$ (eq. 4) | −0.2764 | | | | 0.0008 |
| $n_D$ (eq. 5) | 1.395 | 0 | 0.0023 | | 0.0005 |

Dibutyl ether(1) + octane; $T = 303.15$ K

| | | | | | |
|---|---|---|---|---|---|
| $\Delta\varepsilon_r$ (eq. 4) | −0.2659 | | | | 0.0006 |
| $n_D$ (eq. 5) | 1.393 | 0 | 0.0012 | | 0.0001 |

Dibutyl ether(1) + octane; $T = 308.15$ K

| | | | | | |
|---|---|---|---|---|---|
| $\Delta\varepsilon_r$ (eq .4) | −0.2512 | 0.0060 | | | 0.0005 |

**CAPTION TO FIGURES**

**FIG. 1** (a) Scheme of the measuring cell Agilent 16452A; (b) Scheme of the experimental arrangement

**FIG. 2** $\Delta \varepsilon_r$ at temperature $T$ for the systems investigated. Symbols, experimental values (this work): (■), 1-pentanol(1) + octane(2); (●), 1-pentanol(1) + dibutyl ether(2); (▲), dibutyl ether(1) + octane(2) ($T$ = 298.15 K); (O), 1-pentanol(1) + dibutyl ether(2) at 308.15 K Solid lines, calculations with eq. (4) using the coefficients from Table 4.

**FIG. 3** $n_D$ at 298.15 K for 1-pentanol(1) + octane(2), or + dibutyl ether(2) mixtures. (●) (this work), (O), [29], $\Delta$, [28], octane; (■) (this work), (▲), [30], dibutyl ether.

**FIG. 4** correlation factor, $g_K$, at 298.15 K for the systems investigated: (●) (this work); (O), [28], 1-pentanol(1) + octane(2); (■), 1-pentanol(1) + dibutyl ether(2); (▲), dibutyl ether(1) + octane(2).

**FIG. 5** Molar polarization, $P_m$, at 298.15 K for the systems 1-pentanol(1) + octane(2) (●), or + dibutyl ether(2) (■); and for dibutyl ether(1) + octane(2) (▲).

**FIG. 6** Molar refraction, $R_m$, at 298.15 K for the systems 1-pentanol(1) + octane(2) (●), or + dibutyl ether(2) (■); and for dibutyl ether(1) + octane(2) (▲).

**FIG. 7** $\alpha$ parameter (equation 13) for the systems investigated at 298.15 K. Curves: 8a) 1-pentanol(1) + octane(2); (b) 1-pentanol(1) + dibutyl ether(2); (c) dibutyl ether(1) + octane(2)

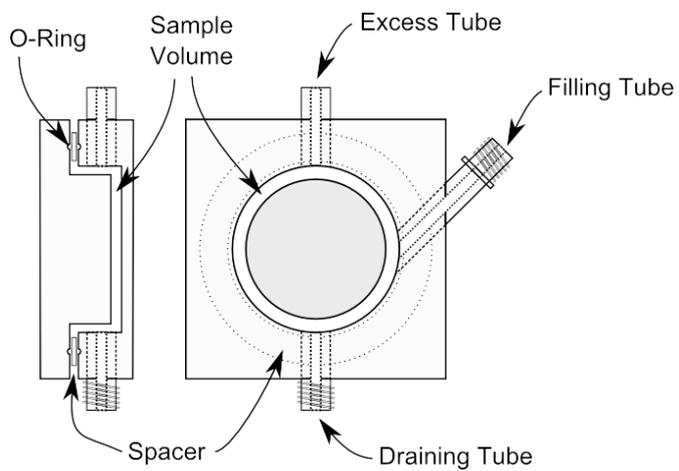

FIGURE 1a

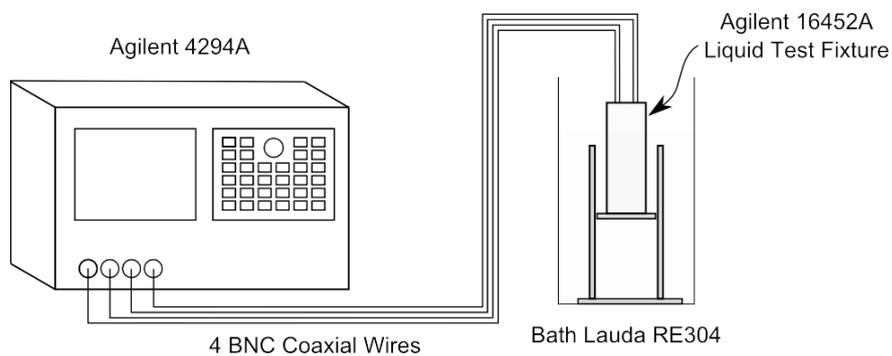

FIGURE 1b

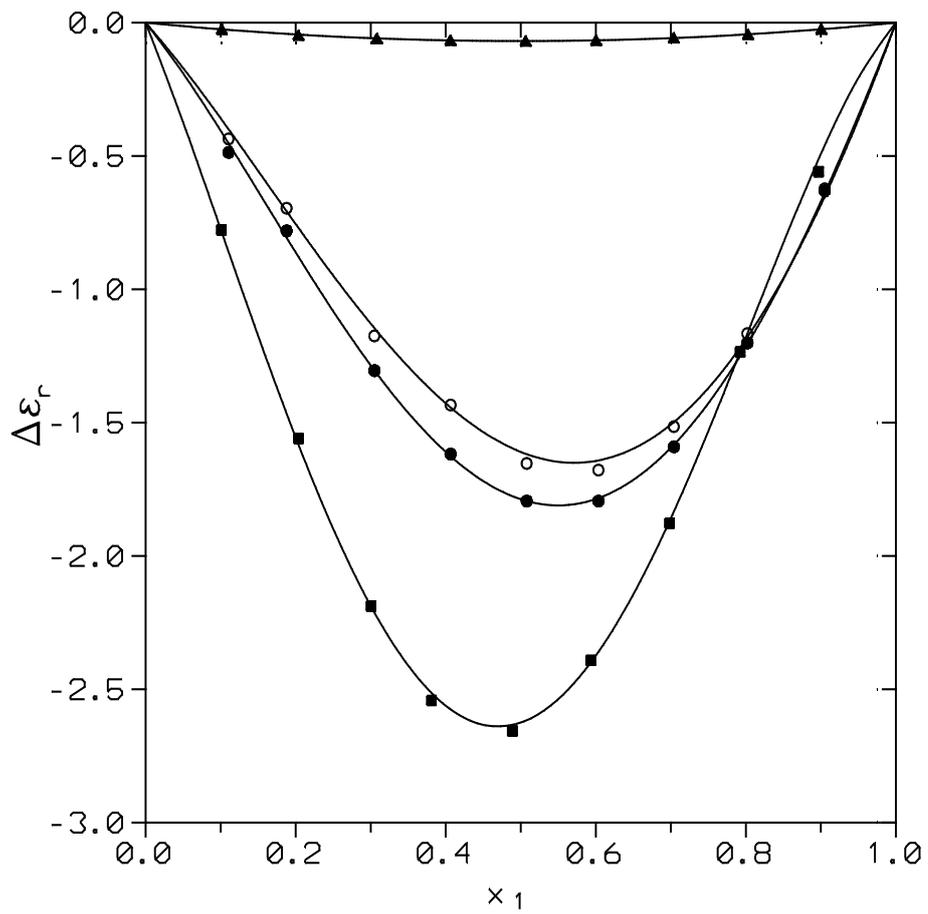

FIGURE 2

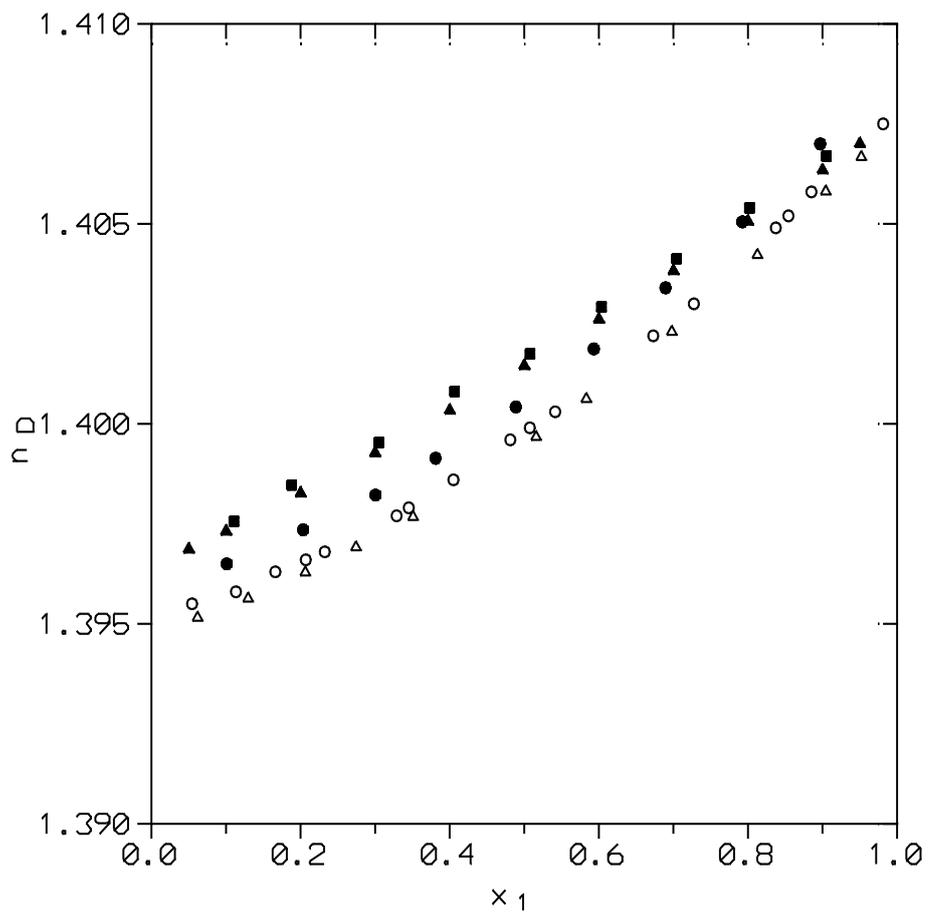

FIGURE 3

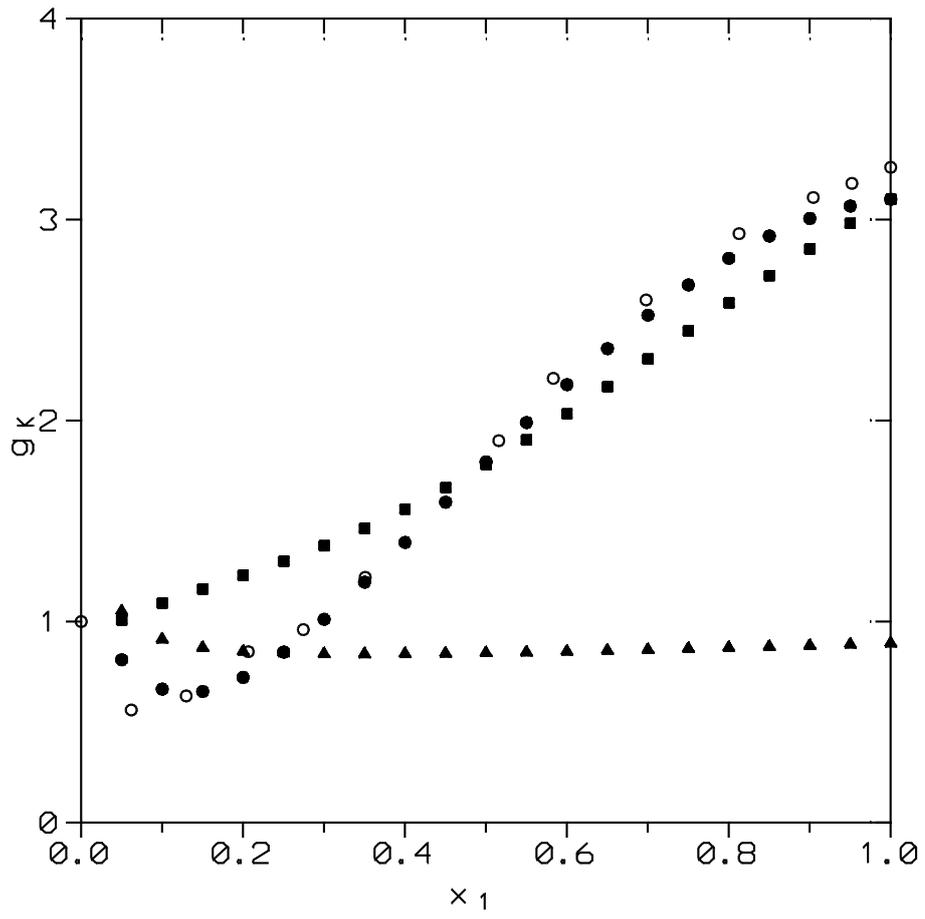

FIGURE 4

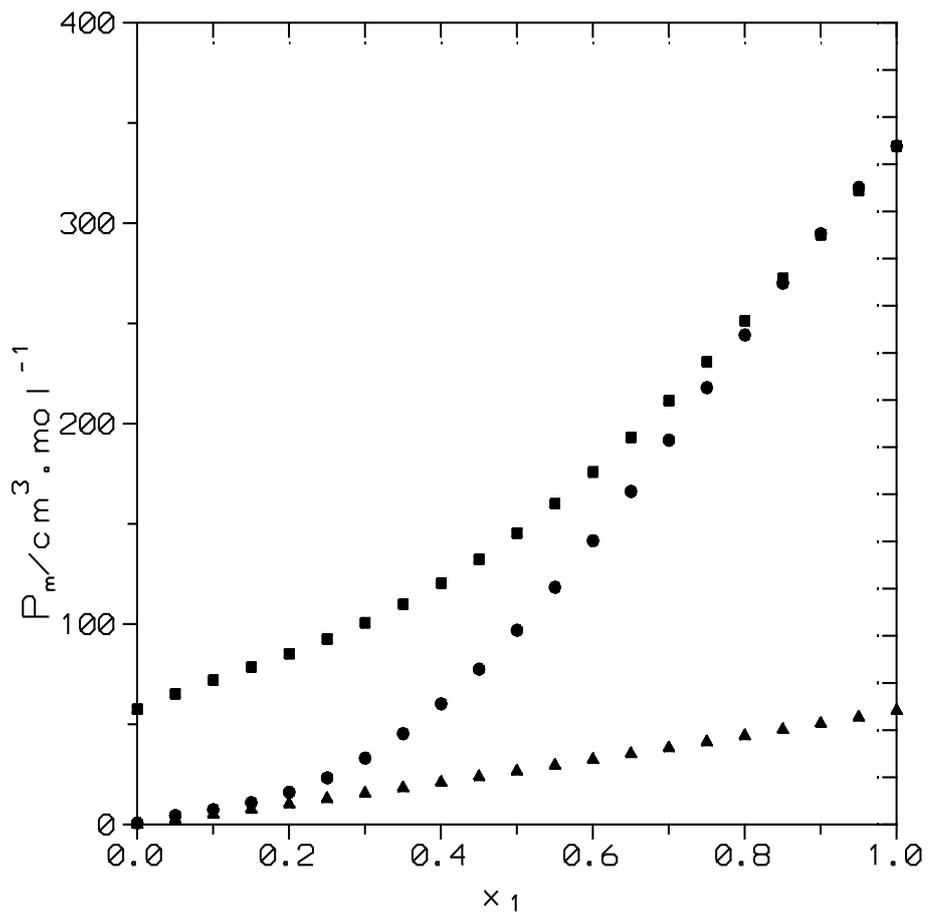

FIGURE 5

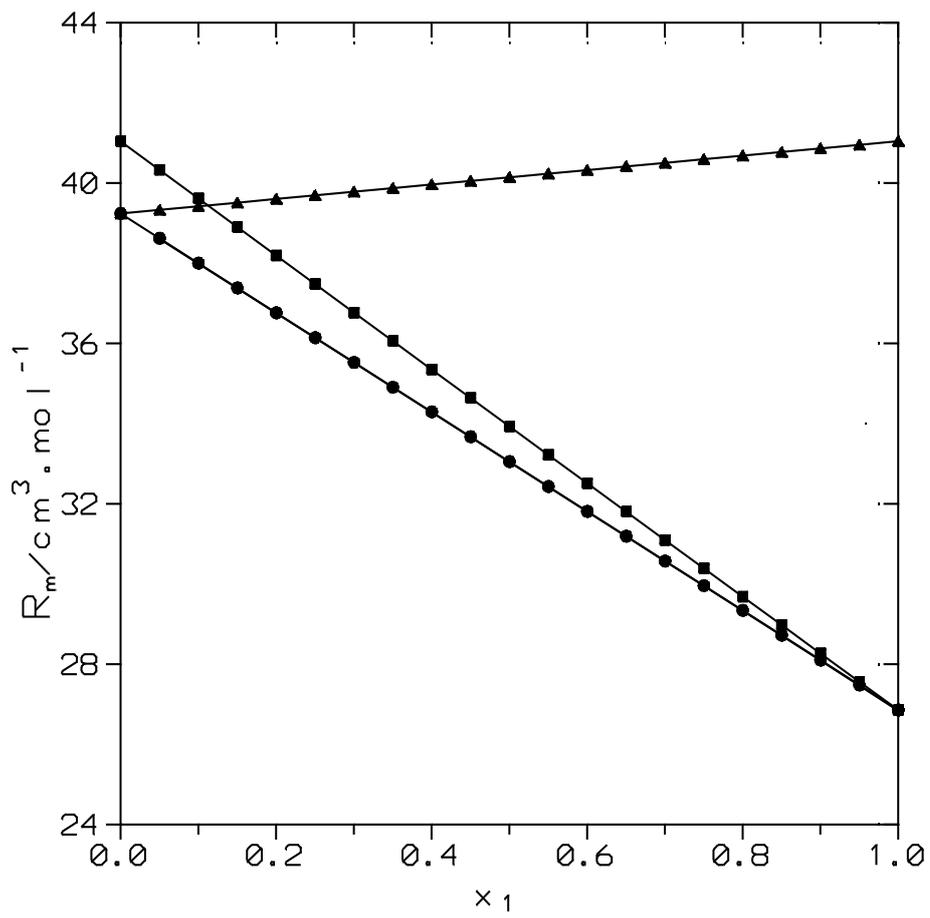

FIGURE 6

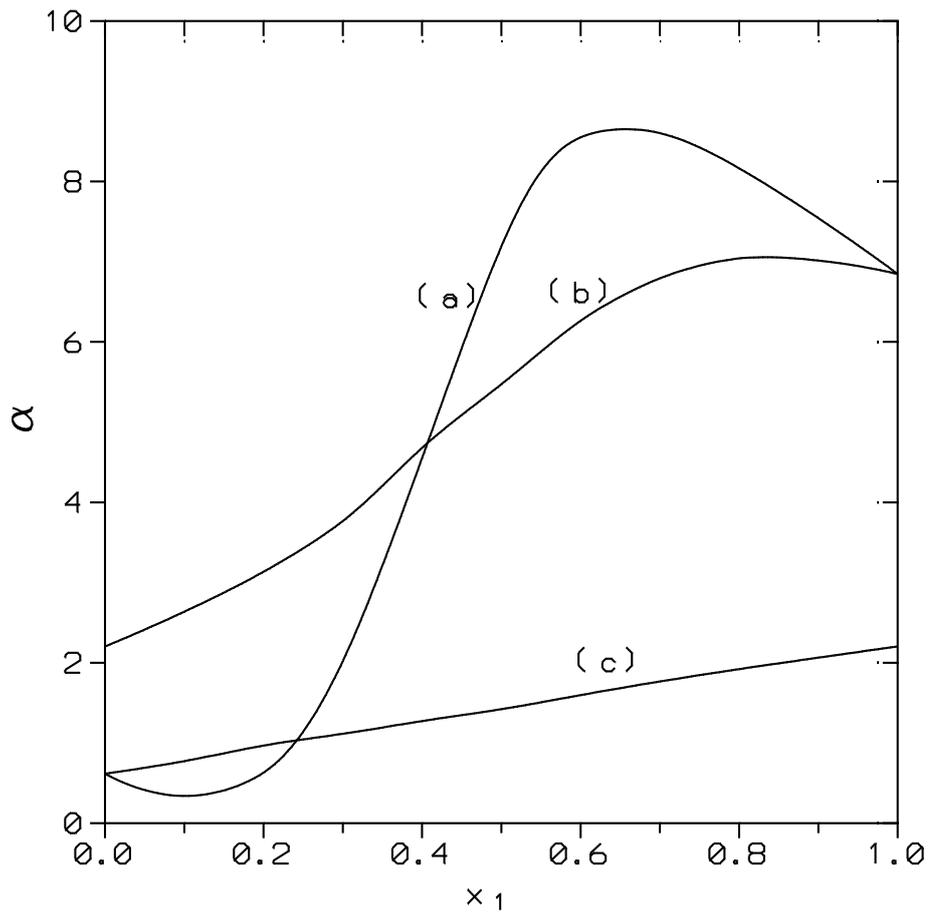

FIGURE 7